\documentclass[conference, comsoc]{IEEEtran}  

\usepackage{cite}
\usepackage{amsmath,amssymb,amsfonts}
\usepackage{graphicx}
\usepackage{graphics}

\usepackage[lofdepth,lotdepth]{subfig}

\usepackage{algorithm}
\usepackage{algorithmic}
\usepackage{multirow}
\usepackage{longtable}
\usepackage{array}
\usepackage{textcomp}
\usepackage{xcolor}
\usepackage{multicol}
\usepackage{amsthm}

\usepackage{microtype}

\def\BibTeX{{\rm B\kern-.05em{\sc i\kern-.025em b}\kern-.08em
		T\kern-.1667em\lower.7ex\hbox{E}\kern-.125emX}}

\begin{document}

\title{Cooperative UAV-Relay based Satellite Aerial Ground Integrated Networks}

  \author{ 
    \IEEEauthorblockN{Bhola\textsuperscript{*}, 
    Yu-Jia Chen\textsuperscript{$\dagger$}, 
    Ashutosh Balakrishnan\textsuperscript{$\ddagger$}, 
    Swades De\textsuperscript{$\ddagger$}, 
    and Li-Chun Wang\textsuperscript{$*$}}
    \IEEEauthorblockA{\textsuperscript{*}Department of Electrical Engineering and Computer Science, National Yang Ming Chiao Tung University, Hsinchu, Taiwan \\ 
    \textsuperscript{$\dagger$}Department of Communication Engineering, National Central University, Taoyuan, Taiwan\\
    \textsuperscript{$\ddagger$}Department of Electrical Engineering and Bharti School of Telecommunications, Indian Institute of Technology Delhi, India}
    %
}

\maketitle

\begin{abstract}
In the post-fifth generation ($5$G) era, escalating user quality of service (QoS) strains terrestrial network capacity, especially in urban areas with dynamic traffic distributions. This paper introduces a novel cooperative unmanned aerial vehicle relay-based deployment (CUD) framework in satellite air-ground integrated networks (SAGIN). The CUD strategy deploys an unmanned aerial vehicle-based relay (UAVr) in an amplify-and-forward (AF) mode to enhance user QoS when terrestrial base stations fall short of network capacity. By combining low earth orbit (LEO) satellite and UAVr signals using cooperative diversity, the CUD framework enhances the signal to noise ratio (SNR) at the user. Comparative evaluations against existing frameworks reveal performance improvements, demonstrating the effectiveness of the CUD framework in addressing the evolving demands of next-generation networks.
\end{abstract}

\section{Introduction }
\label{sec:introduction}

The escalating demand for reliable wireless access, driven by technologies like augmented reality, Internet of things (IoT), and autonomous vehicles, is underscored by Cisco's forecast of $13.1$ billion mobile users and $29.3$ billion Internet-enabled devices by $2023$~\cite{liu2022evolution}. The problem is compounded by the space-time varying dynamic cellular traffic creating temporary hotspots, thereby putting stress on the ground base station \cite{AB_TNSM}.  Meeting this demand necessitates new network planning strategies for beyond fifth generation (B5G) communications to maintain QoS while accommodating dynamic supply needs—bodies like  3rd Generation Partnership Project ($3$GPP) advocate integrating terrestrial with non-terrestrial networks to enhance capacity effectively~\cite{9275613}.
%

%

Recent efforts focus on integrating terrestrial networks with LEO satellites to enhance B$5$G connectivity. LEO satellite networks, characterized by mega-constellations, promise high-speed broadband access. While satellites have traditionally served rural areas well, urban environments present challenges due to masking effects from weather conditions and terrestrial obstacles, significantly attenuating satellite signals~\cite{9755995}.
Deploying unmanned aerial vehicles (UAV-BS) has traditionally been used to inject capacity in terrestrial networks~\cite{9535285}, \cite{AB_ICC}. Rather than deploying UAV-BSs, in this paper, we deploy low-altitude UAV-based relays (UAVr) to overcome the challenges associated with integrating LEO satellites and thus enhance the user QoS. Compared to UAV-BS, the proposed strategy of deploying UAVr is observed to be more energy efficient in addition to meeting the user QoS. 
Compared to terrestrial relays in heterogeneous networks, UAVr offers mobility to the mobile operator in the network. Unlike terrestrial relays, UAVr can also be deployed in rural or disaster scenarios like floods.
%

%
Our paper introduces a cooperative UAVr deployment (CUD) framework for SAGIN communication networks. Operating within a fixed terrestrial area, our CUD framework strategically places UAVr based on traffic patterns to enhance network capacity and user QoS. By employing cooperative diversity (CD), our framework optimizes signal reception at ground users, utilizing both LEO satellite and UAVr links. 
%
%
%

%
Section~\ref{sec: System Model} presents the system model, while 
Section~\ref{Adaptive UAVr Placement and Signal Combining} details the proposed CUD framework. 
Section~\ref{sec: Problem Formulation} formulates a network capacity maximization problem and 
Section~\ref{sec: Problem Formulation} presents an algorithmic solution framework. 
Section~\ref{sec: Simulation Results and Performance Analysis} shows the simulation results and performance analysis results, while 
Section~\ref{Sec: Conclusion} concludes the paper.  

\begin{figure}[!t]
    \centering
    \includegraphics[width=\columnwidth]{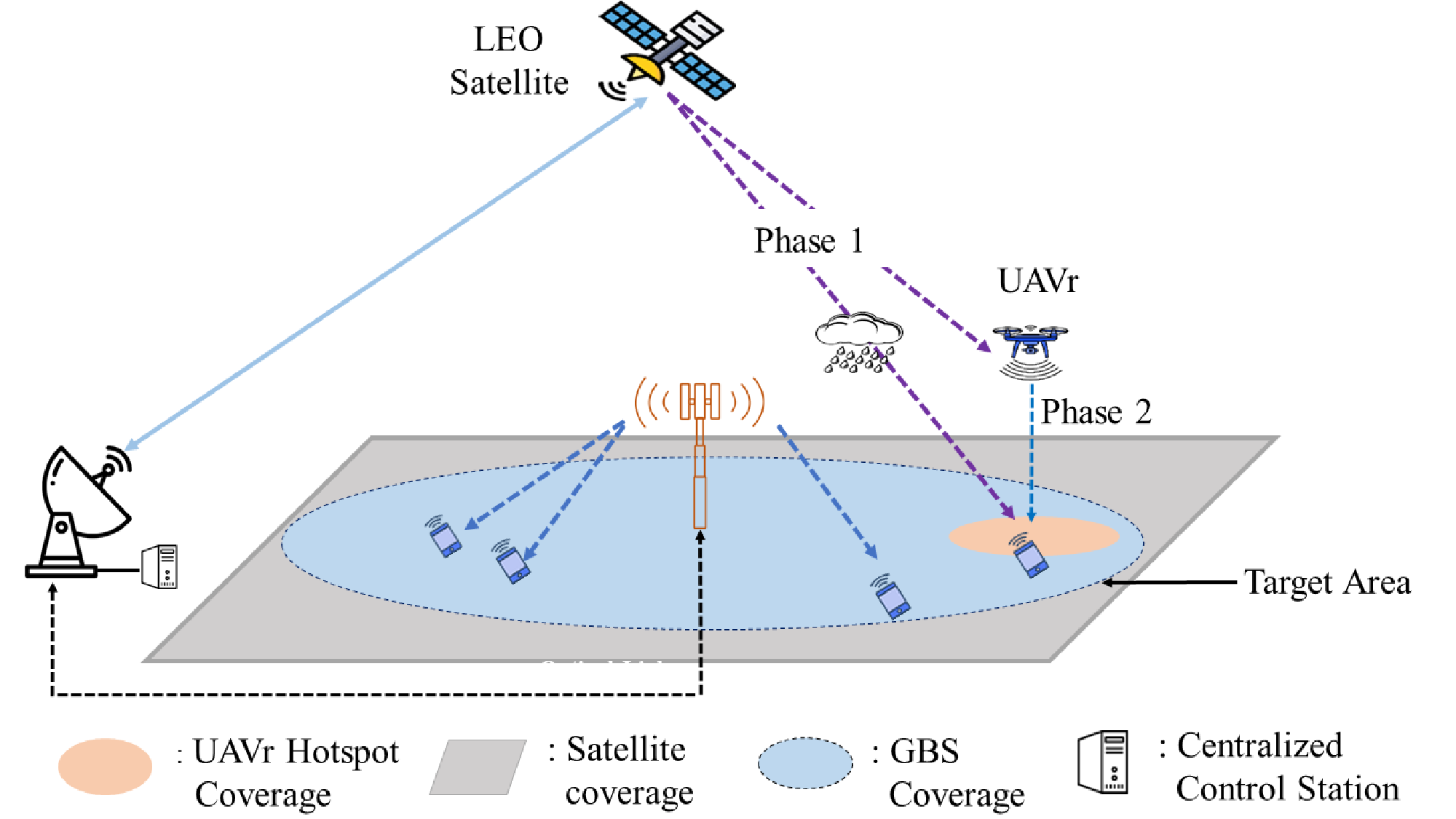}
    \caption{System model overview.}
    \label{fig: system model}
\end{figure}
%
%
%
\section{System Model}
\label{sec: System Model}
  This paper explores downlink resource allocation in a terrestrial network enhanced by UAVr and LEO satellite assistance, as illustrated in Fig.~\ref{fig: system model}. Ground users are represented by a Poisson point process (PPP) within the Ground Base Station (GBS) coverage area, with their activity monitored across discrete time slots indexed by $t$, ensuring slot stability.
  A centralized control station (CCS) oversees the UAVr operation and manages user association and UAVr movements~\cite{9555387}. Mobile user terminals in this network support hybrid communication, connecting with LEO satellites, UAVr, and the GBS~\cite{9385374}.
  %
%
\subsection{Signal Model}
In this subsection, we introduce the signal model for the LEO-assisted UAVr communication framework. The framework comprises a dual-hop cooperative diversity system involving an LEO satellite, UAVr, and ground users, each equipped with $L$ antennas. We assume the user communicates with the UAVr and LEO satellites in the same sub$-6$ Ghz frequency band and with the GBS in different bands at the sub$-6$ Ghz frequency.
 
The channel coefficients linking the satellite to the ground user antenna are denoted by \( {h_0} = \left( \hbar_{i,s_{l}} \right)^{T} \), where \( l \) ranges from 1 to \( L \), $i$ denote user, and $s$ denote the LEO satellite.
The channel coefficients between satellite and UAVr are expressed as \( {h_1} = \left( \hbar_{j,s} \right)^T \), where $j$ denote the UAVr.
Similarly, the channel coefficients between UAVr and ground user antenna are denoted by \( {h_2} = \left( \hbar_{i,j_{l}} \right)^{T} \), where \( l \) varies from 1 to \( L \).~It may be noted that superscripts $(.)^T$ and $(.)^{\dag}$ denote the transpose and transpose conjugate, respectively.
Let's assume that the satellite transmits a signal \( x_{\rm sym} \) with an average power of \( P_s^{\rm tx} \) to the user during the first phase. The signal received at UAVr from the satellite is given by \( r_{s}^{j,s} = h_{1}x_{\rm sym}+n_{1} \), and the signal received at the user from the satellite is \( r_{s}^{i,s} = h_{0}x_{\rm sym}+n_{0} \).
While the satellite remains silent during the second phase, UAVr re-transmits a scaled version of the received satellite signal in fixed-gain AF mode, with the average transmit power from UAVr to the user, $ r_{s}^{i,j} = h_{2}\mathcal{G}(r_{s}^{j,s})+n_{2} = h_{2}\mathcal{G}h_{1}x_{\rm sym}+h_{2}\mathcal{G}n_{1}+n_{2}.$ 
The above dual-hop cooperative communication framework at the user end is written as $\mathbf{r}_{s}^{\text{Tot}} = \mathbb{H} x_{\text{sym}} + \mathbb{N}$, 
where $ \mathbf{r}_{s}^{\text{Tot}} = \left[ \begin{matrix} r_{s}^{i,s} \\ r_{s}^{i,j} \end{matrix} \right], 
  \mathbb{H} = \left[ \begin{matrix} h_{0} \\ h_{2}\mathcal{G}h_{1} \end{matrix} \right], \mathbb{N} = \left[ \begin{matrix} n_{0} \\ h_{2}\mathcal{G}n_{1}+n_{2} \end{matrix} \right],$ 
${\rm n}_{\iota}, \iota = \{0, 1, 2\}$,
Denote $L$ dimensional additive white Gaussian's noise (AWGN) vectors and are modeled as identical and independent random variables, i.e., $n_0, n_1, n_2\thicksim\mathcal{C}\mathcal{N} (0, {\sigma }^2)$, and $\mathcal{G}$ denotes the fixed-gain factor at the UAVr.

 \subsection{Channel Model}  
\label{sec:system}
    %
    This subsection discusses the channel from the LEO satellite to the user and the UAVr, and the UAVr to the user. 
   %
   \subsubsection{ UAVr to User}  

    The spatial $3$D coordinates of a UAVr is denoted as $U^{\mathrm{3}\mathrm{D}}_j\left(t\right)=(x_j\left(t\right),y_j\left(t\right),h^{\rm Ur}_j\left(t\right))$ and the two-dimensional (2D) coordinates $i(t) = (x_i(t), y_i(t))$ of ground users in time slot $t$. Consequently, the horizontal separation between UAVr and ground user can be expressed as, $r_{i,j}(t)=\sqrt{(x_{j}(t)-x_{i})^2+ (y_{j}(t)-y_{i})^2}$.
    Referring to this equation, we can define the Euclidean distance between UAVr and ground user as $   	d_{i,j}\left(t\right)=\sqrt{r^2_{i,j}\left(t\right)+(h^{\rm Ur}_j)^2\left(t\right)}.$
    %
%
    The user's path loss is determined using the air-to-ground channel model from~\cite{al2014optimal}, considering both line of sight (LoS) and non-line of sight (NLoS) scenarios:
    \begin{align}
        PL^\text{LoS}_{d_{i,j}}(t) = 20\log_{10}\left(\frac{4\pi f_c d_{i,j}(t)}{c}\right) + \eta_{{\rm{LoS}}},\\
        PL^\text{NLoS}_{d_{i,j}}(t) = 20\log_{10}\left(\frac{4\pi f_c d_{i,j}(t)}{c}\right) + \eta_{{\rm{NLoS}}}.
    \end{align}
    Here, $\eta_{{\rm{LoS}}}$ and $\eta_{{\rm{NLoS}}}$ represent additional losses due to LoS and NLoS links. The probability of LoS signals from UAVr to the ground user is given by~\cite{al2014optimal}: $ P^\text{LoS}_{d_{i,j}}(t) = \frac{1}{1+a\exp\left(-b\left (\frac{180}{{\rm \pi }}\theta_{i,j}-a\right)\right)}, $
    where $\theta_{i,j}=\tan^{-1}\left(\frac{h^{\rm Ur}_j(t)}{r_{i,j}(t)}\right)$ and $a$, $b$ are environmental factors. The probability of NLoS signals is $ P^\text{NLoS}_{d_{i,j}} (t) = 1-P^\text{LoS}_{d_{i,j}}(t) $.
    %
    The channel gain $\hbar_{i,j}$ between the UAVr and the user is  ~\cite{10164260}
    \begin{equation}
    \label{eq: gain_UAV_user}
    \resizebox{\hsize}{!}{$
        \hbar_{i,j}(t) =  {\mathfrak{g}_{i,j}}{{\left( \frac{4\pi {{f}_{c}}{{d}_{i,j}(t)}}{c} \right)}^{-\frac{{{\alpha_{exp} }^{\left( 3 \right)}}}{2}}}  \times {{10}^{-{\frac{P_{d_{i,j}}^{{\rm{LoS}}}(t)\times PL_{d_{i,j}}^{{\rm {LoS}}}(t)+P_{d_{i,j}}^{ {\rm{NLoS}}}(t)\times PL_{d_{i,j}}^{{\rm{NLoS}}}(t)}{20}}}},
        $}
    \end{equation}
    Here, $d_{i,j}$ represents the distance between a UAVr and a user, $\alpha_{exp}^{(3)}$ denotes the path loss exponent from the UAVr to the user, and $\mathfrak{g}_{i,j}$ signifies the small-scale fading component of the link channel between the UAVr and the user.
    The average path loss of the signal from UAVr to the ground user is 
    \begin{align}
        &PL_{d_{i,j}}^{{\rm{Avg}}}(t) = P_{d_{i,j}}^{{\rm{LoS}}}(t)\times PL_{d_{i,j}}^{{\rm {LoS}}}(t) + P_{d_{i,j}}^{ {\rm{NLoS}}}(t)\times PL_{d_{i,j}}^{{\rm{NLoS}}}(t) \notag \\
        & = \frac{A}{{1+a\exp\left({-b\left[{{\frac{180}{{\rm \pi }}\theta_{i,j}}-a}\right]}\right)}} + 20{\log_{10}}\left ({{d_{i,j}(t)}}\right) + \beta,
    \label{Eq:09} 
    \end{align}
    where $\beta=20\log_{10}\left(\frac{4\pi f_c}{c}\right) +\eta_{{\rm{NLoS}}}$ and ${A = \eta_{{\rm{LoS}}}}-{\eta_{{\rm{NLoS}}}}$. 

    Let \( p_{i,j} \) denote the minimum required transmission power to send a signal from the UAVr to user. Hence, the SNR associated with UAVr is ${\gamma }_{i,j}(t)=p_{i,j}.{\left\| {\hbar_{i,j}(t)} \right\|}^{2} \slash {B_{i,j}{{\sigma }^2}}$.
    The hovering power of the UAVr is computed as $
	p_{j}^{\rm Hov}=p_{0}(1+\Delta)e^{{{\rm \varepsilon}h^{\rm Ur}_{j}\mathord{\left/{\vphantom{{\rm \varepsilon}h^{\rm Ur}_j2}}\right.\kern- \nulldelimiterspace}2}}.$
    Here, $p_{0}$ denotes the power the active UAVr utilizes during hovering. $\Delta$ and $\epsilon$ are constants, while $h^{\rm Ur}_j$ indicates the altitude of the UAVr \cite{9946428}. The hovering altitude of the UAVr, corresponding to its hovering power, is $h^{\rm Ur}_j = 2 \slash {\epsilon}{\rm{ln}}\frac{{p_j^{{\rm{Hov}}}} }{{{p_0}(1+\Delta)}}$.
    The total power (communication and hovering) consumption of  UAVr is
	  $p_{j}^{{\rm Total}}(t)= p_{i,j}(t) + p_{j}^{\rm Hov}(t).$

    \subsubsection{ LEO satellite to UAVr and User }  

   We consider mobility of LEO satellites in assessing system performance, which are non-geostationary. The link between a UAVr and a LEO satellite is encapsulated through a visibility parameter $V_{j,s}(t)$, evaluated at  time slot $t$ as outlined in~\cite{9775682}.
     \begin{equation}
     \label{eq: Satellite_visibility}
        {V }_{j,s}(t)=\left\{ \begin{array}{c} 1\ \mathrm{\ }\mathrm{\ }\mathrm{if}\ {\mathrm{cos} \left(\frac{\mathrm{2}\mathrm{\piup }\mathrm{t}}{T_{\mathrm{s}}}-{\theta }_p\right)\ge \ }\frac{R^2_E+{r_{\rm EC}^2}-d^2_{\mathrm{SR}}}{2R_E{r_{\rm EC}}} \\ \ 0\ \ \ \ \ \ \ \ \ \ \ \ \ \ \ \ \ \ \mathrm{otherwise}.  \ \ \ \ \ \ \ \ \ \ \ \end{array} \right.
    \end{equation}    
    %
    Sophisticated handover (HO) schemes, like guaranteed and prioritized HO, ensure seamless transitions between LEO satellites and UAVr, maintaining data transmission reliability. 
    The channel gains from satellite to UAVr are described by the shadowed-Rician fading (SRF) model: ${\hbar}_{j,s} = \sqrt{{{\mathfrak{g}^{\rm {avg}}}}d_{j,s}^{-\alpha_{exp}^2}}$, where $\mathfrak{g}^{\rm {avg}}$ represents average channel gain, $d_{j,s}$ is the satellite-UAVr distance, and $\alpha_{exp}^2$ denotes the path loss exponent. The SRF component ${\mathfrak{g}^{\rm {avg}}}\thicksim$ SR ${\left( {{\wp}},{{\Im}},{{\varnothing}} \right)}$ includes direct signal average power $\wp$, half average power of the scatter portion $\Im$, and Nakagami-m fading component $\varnothing$.

    The instantaneous SNR for each communication link is determined based on the LEO satellite visibility criterion~\eqref{eq: Satellite_visibility}.
   \begin{equation}
   \label{eq: Satellite_UAV}
        {{\gamma_{j,s} }}(t)={V }_{j,s}(t){{{P_s^{\rm tx}}}{{\left\| {\hbar_{j,s}} \right\|}^{2}}}/{{{\sigma }^2}}\;.
   \end{equation}
    Similarly, the channel gain from satellite to user is expressed as $\hbar_{i,s} = \sqrt{{{\mathfrak{g}^{\rm avg}}}d_{i,s}^{-\alpha_{exp}^2}}$ 
    with the SNR being ${{\gamma_{i,s} }}(t)={V }_{i,s}(t){{{P_s^{\rm tx}}}{{\left\| {\hbar_{i,s}} \right\|}^{2}}}/{({{\sigma }^2})}.$
    Here, $ P_{{s}}^{{\rm{tx}}} $ denotes the average transmission power from the satellite to both the UAVr and user.

    
\subsubsection{ GBS to User }  
   %
   It is presupposed that the user undergoes independent Rayleigh fading while connected to the GBS. At each time slot $t$, the channel coefficient is $
        \hbar_{i,\scriptstyle \textit{G}}\left(t\right)= {\mathfrak{g}_{i,\scriptstyle \textit{G}}\left(r_{i,\scriptstyle \textit{G}}\left(t\right)\right)}^{-\alpha_{exp} }.$
    %
   Here, $\mathfrak{g}_{i,\scriptstyle \textit{G}}$ denotes the small-scale Rayleigh fading gain, following a complex Gaussian distribution, i.e., $g_{i,\scriptstyle \textit{G}}\sim N (0,1)$. $r_{i,\scriptstyle \textit{G}}$ represents the distance between the user and the GBS. Hence, the SNR for users associated with GBS is
$    {\gamma }_{i,\scriptstyle \textit{G}}\left(t\right)=P_{{\scriptstyle \textit{G}}}^{{\rm{tr}}}{\left\|\hbar_{i,\scriptstyle \textit{G}}\left(t\right)\right\|}^2 \slash B_{i,\scriptstyle \textit{G}}{\sigma }^2. $
   Here, $B_{i, \scriptstyle \textit{G}}$ is the bandwidth allocated to the user, and the GBS transmits at a fixed power \(P_{\scriptstyle \textit{G}}^{\rm{tr}}\) for terrestrial communications. UAVr/Satellite links are utilized only when the GBS cannot serve more users to maximize network capacity. Hence, we introduce a binary indicator \(\delta_{i, \scriptstyle \textit{G}}(t)\) for a user who meets the SNR and load conditions under GBS coverage at time \(t\):
\begin{equation}
    \delta_{i,\scriptstyle \textit{G}} = 
    \begin{cases}
      1, & \text{if} \ \left({\gamma_{i,\scriptstyle \textit{G}}} \ge {\gamma}_{\mathrm{th}}\right) \wedge (|\Omega_{\scriptstyle \textit{G}}| \leq \omega_{\scriptstyle \textit{G}}^{\text{max}}) \\
      0, & \text{otherwise.}
    \end{cases}
    \label{Eq:gbs_indicator}
\end{equation}
  Here, ${\gamma }_{\mathrm{th}}$ is defined as the predefined SNR threshold for successful signal transmission, $\Omega_{\scriptstyle \textit{G}}$ and $\omega_{\scriptstyle \textit{G}}^{\text{max}}$ show the current user associated with GBS and the maximum user association capacity of GBS.
  Therefore, the achievable data rate for users connected to a GBS is
     $   c_{i, \scriptstyle \textit{G}}(t) = B_{i,\scriptstyle \textit{G}}\log_2(1+\gamma_{i,\scriptstyle \textit{G}}(t)) \delta_{i,\scriptstyle \textit{G}}(t).$
    %
    %

\section{ UAV Relay Placement and Signal Combining}
    \label{Adaptive UAVr Placement and Signal Combining}
   When the ground base station falls short of meeting user QoS, our CUD method deploys UAVr to the hotspots, leveraging LEO satellites to serve users as explained in this section.

\subsection{UAVr Coverage Analysis}
    \label{UAVr coverage analysis}
        
    A user falls within UAVr coverage if its distance from the center of the UAVr coverage region, $r_{i,j}(t)$, satisfies the condition, $
        \delta_{i,j}r^2_{i,j}(t) \le R^2_j(t).$  
Here $R_j$ denoted the UAVr radius. We adapt it following~\cite{7918510}, resulting in:
    \begin{align}
        r^2_{i,j}(t) \le R^2_j(t) + M(1-\delta_{i,j}).
        \label{Eq:26}
    \end{align}
    $M$ is a large constant, indicating significant distance when $\delta_{i,j}=0$. The user is associated with $\delta_{i,j} = 1$, where $\delta_{i,j} \in\{0,1\}$ serves as an indicator function denoting user association with an access point represented as:
    \begin{equation}
        \delta_{i,j} = 
        \begin{cases}
            1, & \text{if } r^2_{i,j}(t) \le R^2_j(t) \\
            0, & \text{otherwise.}
        \end{cases}
    \label{eq: Indicater_UAV_user}
    \end{equation}
   
    
    The following subsection discusses cooperative diversity (CD) based signal combining employed within the proposed CUD framework. 
    %

\subsection{Cooperative Diversity at User}
    %
    Cooperative communication via LEO satellite-UAVr occurs over two phases: Phase I and Phase II.
    In Phase I, the LEO satellite sends a signal to both UAVr and ground user simultaneously. UAVr then employs AF protocol in Phase II to relay the signal to the user while the satellite remains silent.
    Assuming perfect synchronization between signals from the satellite and UAVr at the user, operating in time division duplex mode, our framework prioritizes efficient signal relaying using AF for its simplicity and lower complexity~\cite{labrador2009approach}.
    Then, using the CD at receiver with the weighting vector ${\mathbf{w}^{\dag}}$, we can express the total output as 
        $\mathbf{w}^{\dagger} \mathbf{r}_{s}^{\text{Tot}}=
        \mathbf{w}^{\dagger} \mathbb{H} \mathbf{x}_{\text{sym}} + \mathbf{w}^{\dagger} \mathbb{N}.$
    In the presence of complete channel state information (CSI) at the destination, the instantaneous SNR at the user is
       $ {\gamma_{\text{\rm AF}}}(\mathbf{w})=\left({P_{s}^{\text{tx}}}  \mathbf{w}^{\dagger} \mathbb{H}\mathbb{H}^{\dagger} \mathbf{w}\right) \slash \left({{\mathbf{w}}^{\dagger }} {{\mathscr{R}}_{\text{n}}} \mathbf{w}\right)$ 
    %
    where, $\mathscr{R}_{\rm n}= \mathbb{E}\{ \mathbb{N} \mathbb{N}^{\dagger} \}$, and ${P_{s}^{\text{tx}}}=\mathbb{E} \left\{ \mathbf{x}_{\text{sym}} \mathbf{x}_{\text{sym}}^{\dagger} \right\}$.
     Taking the derivative of $ {\gamma_{\text{\rm AF}}}(\mathbf{w})$ w.r.t.  the weight vector $\mathbf{w}$~\cite{holter2002optimal}, we get the optimal weight vector in a dual-hop cooperative communication system is~$\mathbf{w}_{\rm opt} = c_{r} \mathscr{R}_{n}^{-1} \mathbb{H}.$
    where $c_{r}$ denotes an arbitrary constant for any \( c_{r} \ne 0 \). 
    Using the optimal weight vector, we can obtain the maximum signal~\cite{4543059} in a dual-hop AF cooperative communication is
    \begin{align}
        \gamma_{\text{AF, max}}^{\text{CD}}(\mathbf{w}_{\text{opt}}) 
        &= \gamma_{i,s} + \frac{\gamma_{j,s}\gamma_{i,j}}{\gamma_{i,j} + \varsigma} 
    \label{eq: SNR_max_1}
    \end{align}
    where $\varsigma =\frac{p_{i,j}}{{{\sigma }^{2}}{\mathcal{G}^{2}}}.$ To represent whether the ground user is associated with the UAVr or not, the indicator function $\delta_{i,j}$ is modified as follows to incorporate the user QoS as well as coverage constraints. 
   \begin{equation}
        \delta_{i,j} = 
        \begin{cases}
            1, \ &\text{if,} \ \left(\gamma_{\text{AF, max}}^{\text{CD}}\left(t\right)\ge {\gamma }_{\mathrm{th}}\right) \mathrm{\wedge } \ ( r^2_{i,j}(t)\le R^2_j(t)) \\
            0, \ &\text{otherwise.} 
        \end{cases}
        \label{Eq:27}
    \end{equation} 
     
    It is assumed that each user can only connect to UAVr at a time, and such a constraint is written as $\sum_{j=1}^{N_{\rm U}}\delta_{i,j}=1$.


    In a time-division cooperative communication scheme, the transmission process is split into two slots: one for the LEO satellite and one for the UAVr. This halves the effective bandwidth for each phase, so the capacity per slot is divided by two to match the overall throughput of a non-cooperative system.
    %
    Thus, the achievable data rate of the user associated with UAVr and LEO Satellite is obtained from the Shannon theorem, expressed as
    \begin{align}
	    {c}^{ \mathrm{CD}}_{{\rm AF}}(t) = \frac{B_{i,j}}{2}\log_2 \left(1+  \gamma_{\text{AF, max}}^{\text{CD}}(t) \right).\delta_{i,j}(t). 
	    \label{Eq:14} 
    \end{align}
    Here, $B_{i,j}$ is the allocated bandwidth (MHz) of the downlink connection from the UAVr to the user and the LEO satellite to the user. 
    According to~\eqref{Eq:14}, the data transmission rate achievable by users associated with LEO satellites through UAVr is 
	 $   {C_{j}}(t)=\sum_{i\in ~{\Omega_j},\forall {i}\in\{1,2,\dots,N_{\rm U}\}} {c}^{ \mathrm{CD}}_{{\rm AF}}(t)$.
    %
    where $\Omega_j=\text{max}(0, |\Omega_{\rm Tot}|-\omega_{G}^{\text{max}}) $ is the set of users associated with the collaboration with UAVr/LEO Satellite, $\Omega_{\rm Tot}$ is the total set of users in the system, $\omega_{G}^{\text{max}}$ is the maximum user association capacity of GBS, and  $N_{\rm U}$ is the number of users served by UAVr/LEO Satellite collaboratively. The total GBS capacity is computed as 
    %
        ${C_{\scriptstyle \textit{G}}}(t)=\sum_{i \in ~{\Omega_G},\forall {i}\in\{1,2,\dots,N_{\scriptstyle \textit{G}}\}}{c_{i,\scriptstyle \textit{G}}}(t)$.
    $\Omega_{\scriptstyle \textit{G}}$ is the set of users associated with the GBS, and  $N_{\rm \scriptstyle \textit{G}}$ is the users served by the GBS.
    The total energy efficiency is dervied as
    \begin{align}
        E^{\rm Tot}(t)= \frac{C_j(t) + C_G (t) }{(p_{j}^{\rm Total}(t)+P_{\scriptstyle \textit{G}}^{\rm{tr}}(t))} 
    \label{eq: energy_efficiency}
    \end{align}
    %
%
\section{ Problem Formulation and Proposed Solution} 
\label{sec: Problem Formulation}
We aim to increase the network's total capacity through the proposed CUD approach involving the adaptive deployment of UAVr at the desired location, followed by signal combining from the LEO satellite and UAVr. 

\subsection{ Problem Formulation }
The minimum data rate and user association specifications constrain the problem formulated below. 
The problem formulation along with constraints is defined as:
\begin{equation}
    \max_{\delta_{i,j} (t), p_{i,j} (t)} (C_{j}(t) + C_G(t)) \; 
    \label{Eq:obj_P1}
\end{equation}
\vspace{-7mm}
\begin{subequations}
    \renewcommand{\theequation}{\theparentequation\alph{equation}}
    \begin{align}
    &\text{subject to}~\eqref{eq: SNR_max_1},~\eqref{Eq:27},~\eqref{Eq:14},\nonumber\\
        &(C1.) \ 0  \leq p_{i,j} \leq p_{\max}, \ (C2.) \ 0  \leq |\Omega_{j}| \leq \omega_{j}^{\mathrm{max}}, 
        \label{Eq:31} \\
        &(C3) \ 0 \leq |\Omega_{G}| \leq \omega_{G}^{\mathrm{max}}, \ (C4) \ ~r^2_{i,j}(t)\le {R^2_{j}(t)}, \notag\\
        &(C5) \ ~{\delta}_{i,j}\in \{ 0,1 \},
        \ \forall {i}\in {\Omega_j}, \ (C6) \  ~{\delta}_{i ,G} = 1 - \delta_{i,j}, \ \forall i.
        \label{Eq:new_constraint} 
    \end{align}
\end{subequations}
    
   Constraints (C1) limit transmission power for UAVr. Constraints (C2) and (C3) restrict user associations with UAVr and GBS. Candidate configurations, including UAVr locations $U_j^{\rm 2D}$ and coverage radius $R_j$, derive from relationships in (C4), obtained from (C5) and Constraint (C6) ensures each user is associated with either the GBS or the UAVr-satellite. For user, $\delta_{i,G} = 1$ indicates GBS association, while $\delta_{i,j} = 1$ indicates UAVr association. 

   %
   Problem~\eqref{Eq:obj_P1} presents a non-convex mixed-integer programming challenge due to the complex mathematical formulation of successful transmission and the integer association constraint~\eqref{Eq:27}. Solving such problems typically falls outside the realm of polynomial computational complexity. 
   However, optimizing transmission power with a fixed user association becomes significantly more manageable~\cite{7070670}. Therefore, we introduce a CUD-based approach to optimize the UAVr transmission power $p_{i,j}$ by fixing the association variable ${\delta}_{i,j}$.
   %
   
%
\begin{algorithm}[!ht]
    \caption{Cooperative signal combining based UAVr deployment framework} 
    \label{alg: CUD}
    \small
    \textbf{Input:} $\omega^{\mathrm{max}}_{j}, \omega^{\mathrm{max}}_{G}, {\Omega}, \Omega^{\mathrm{Ex}}_{G}, {\textit{D}^{\rm th}_{j}}$ 
    \begin{algorithmic}[1]
        \STATE Consider two types of hotspots: 
        \label{alg: CUD:01}
        \STATE a) Density-based: ${\omega^{\mathrm{\rm max}}_{j} = D^{\mathrm{\rm th}}_{j}*\pi*R^{\mathrm{\rm 2}}_{j}}$
        \STATE b) User-based: ${\Omega}^{\mathrm{\rm {Ex}}}_{G}$=$(\Omega_{G}-\omega^{\mathrm{\rm max}}_{G})$
        \FOR{$j=1$} 
        \label{alg: CUD:02}
        \WHILE{$( \omega^{\mathrm{max}}_{G}>0 \vee \Omega^{\mathrm{Ex}}_{j}>0 ) \wedge V_{j,s}=1 \wedge \gamma_{i,s} < \gamma_{\mathrm{th}}$}
        \label{alg: CUD:03}
        \STATE Select hotspot user $i^{*}$ from $\Omega_{G}$; then send UAVr $U^{*}_{j}$ to serve $i^{*}$;
        \label{alg: CUD:04}
        \FOR{$j=1$}
        \label{alg: CUD:05}
        \FOR{$i={i}$ to $|\bf\Omega_j|$}
        \label{alg: CUD:06}
        \STATE Update distance: ${\rm\bf d}\left( {i}^{*},j^{*} \right)= \sqrt{{\rm\bf r}^2\left( {i}^{*},j^{*} \right)+{h^{\rm Ur}_j}}$
        \label{alg: CUD:07}
        \STATE Determine average path loss and minimum power for SNR: $PL^{\mathrm{Avg}}_{h^{\mathrm{Ur}}_j,r_{{i}^{*},j^{*}}}$, $p^{\mathrm{min}}_{{i}^{*},j^{*}}$
        \label{alg: CUD:08}
        \STATE Find optimal power $p_{{i}^{*},j^{*}}$\cite{9946428} to maximize Eq.\ref{Eq:obj_P1}
        \label{alg: CUD:09}
        \IF{$p^{\mathrm{min}}_{{i}^{*},j^{*}} \leq {p_{\max}}$} 
        \label{alg: CUD:10}
        \IF{$p_{{i}^{*},j^{*}} \leq p^{\mathrm{min}}_{{i}^{*},j^{*}}$} 
        \label{alg: CUD:11}
        \STATE $p_{{i}^{*},j^{*}} = p^{\mathrm{min}}_{{i}^{*},j^{*}}$
        \label{alg: CUD:12}
        \ELSIF{$p_{{i}^{*},j^{*}} \geq {p_{\max}}$} 
        \label{alg: CUD:13}
        \STATE $p_{{i}^{*},j^{*}} = {p_{\max}}$ 
        \label{alg: CUD:14}
        \ENDIF 
        \ELSE 
        \label{alg: CUD:16}
        \STATE $p_{{i}^{*},j^{*}} = {p_{\max}}$
        \label{alg: CUD:17}
        \ENDIF  \label{alg: CUD:18}
        \STATE Transmit power: ${\rm\bf P}\left( i^{*},j^{*} \right)={p_{{i}^{*},j^{*}}}$  \label{alg: CUD:19}
        \ENDFOR \label{alg: CUD:20}
        \ENDFOR  \label{alg: CUD:21}
        \ENDWHILE \label{alg: CUD:22}
        \ENDFOR \label{alg: CUD:23} \\
        \STATE Send ${\rm\bf\Omega}$ and ${\rm\bf P}$ to all UAVr
        \label{alg: CUD:24}
    \end{algorithmic}
\end{algorithm}

\begin{table}[!ht]
	\centering
	\caption{Simulation Parameters}
	\begin{tabular}{|p{3.1cm}|l|l|}
		\hline
		\textbf{Parameter} & \textbf{Symbol} & \textbf{Value}
		\label{Simulation Parameter}\\\hline 
    Environmental parameters & $ a,b,\eta_{\rm LoS},\eta_{\rm NLoS} $ & $ 9.61,0.16,1,20 $ \\ \hline 
    Maximum path loss & $PL_{d_{i,j}}^{\rm{max}}$  & $119$ dB \\ \hline 
   LEO Satellite Altitude & $h_s$ & $500$ km \\ \hline 
    UAVr Altitude and Radius & $ h^{\rm Ur}_j$, $R_{j}$ & $30$m, $100$m \\ \hline
    UAVr and GBS Number & $U_{j},G$ & $1,1$ \\ \hline
    Allocated Bandwidth & $B_{i,j}$ & $20$ MHz \\ \hline 
    Noise Power  &$\sigma^2$ & $-174$ dBm/Hz \\\hline 
    %
    %
    Total Number of Users & $i$ & $200$ \\ \hline
    Max Associations at UAVr & ${{\omega^{\mathrm{\rm max}}_{j}}}$ &     $100$ \\ \hline 
    Max Associations at GBS & ${{\omega^{\mathrm{\rm max}}_{G}}}$ & $100$ \\ \hline
    SNR Threshold & $\gamma_{\rm th}$ & $3$ dB  \\ \hline
    Max Transmission Power & $p_{\rm max}$ & $20$ dBm  \\ \hline
	\end{tabular}
\end{table}

%
%
\subsection{Proposed Solution}


Algorithm~\ref{alg: CUD} presents the pseudo-code for the main procedure of the proposed CUD. The CCS executes this procedure. Key lines within the CUD procedure are outlined below:

\begin{itemize}
    \item The initial phase (Line~\ref{alg: CUD:01}) involves preparing temporary matrices by the CCS, storing information for subsequent re-association and optimization.
    
    \item Line~\ref{alg: CUD:02} systematically examines for overload conditions using a for-loop. Line~\ref{alg: CUD:03} monitors excess users and initiates re-association and transmission power optimization.
    
    
    \item Line~\ref{alg: CUD:04} selects ${i}^{*}$ from $\Omega_{\rm G}$, determining the sector of the user-generated hotspot. Further, Lines~\ref{alg: CUD:05} to~\ref{alg: CUD:23} optimizes transmit power for each UAVr with CCS assistance.
    
    
    \item Line~\ref{alg: CUD:07} computes the Euclidean distance between the UAVr and ${i}$-th user while  Line~\ref{alg: CUD:08} updates the path loss.
    
    
    \item Line~\ref{alg: CUD:09} finds the optimal transmit power to maximize~\eqref{Eq:obj_P1}.  Lines~\ref{alg: CUD:10} to~\ref{alg: CUD:18} check transmit power constraints and commit updated values.
    
    
    \item Finally, the CCS sends updated parameter sets, $\rm\bf\Omega$ and $\rm\bf P$, to all UAVr for deployment updates.
\end{itemize}
 

%
\section{Results and Discussion}
\label{sec: Simulation Results and Performance Analysis}
This section evaluates the problem \eqref{Eq:obj_P1} based on key performance metrics: total capacity and energy efficiency. A fixed adaptive UAVr deployment within a defined area and varying user generation (temporary hotspots) form the basis of our simulations, validating the effectiveness of the proposed CUD approach. The simulations are performed in MATLAB R2020b for an urban setting, with parameter values detailed in Table \ref{Simulation Parameter}.
Furthermore, comparisons are drawn with equal gain combining-based SAGIN (EGC-SAGIN), LEO-GBS, and GBS-Only frameworks. 


%

\subsection{Simulation results}
In Fig. 2(a), we compare the performance of CUD with state-of-the-art frameworks across varying excess user counts under the GBS coverage. CUD consistently outperforms other setups in total network capacity, achieving gains of up to $0.034$\% ($13100$ bps) compared to the GBS-only framework. Regarding EGC-SAGIN and LEO-GBS frameworks, CUD achieves enhancements of up to $0.017$\% ($6600$ bps) and $0.024$\% ($9100$ bps), respectively.
The capacity improvements by CUD, facilitated by strategic UAVr deployment and CD-based signal combining, underscore its efficacy over existing non-SAGIN configurations. These findings highlight the rationale for operators to consider SAGIN-based network architectures.
       %

\begin{figure}[t]
    \centering 
        \subfloat[]{
    \includegraphics[width = 0.25\textwidth]{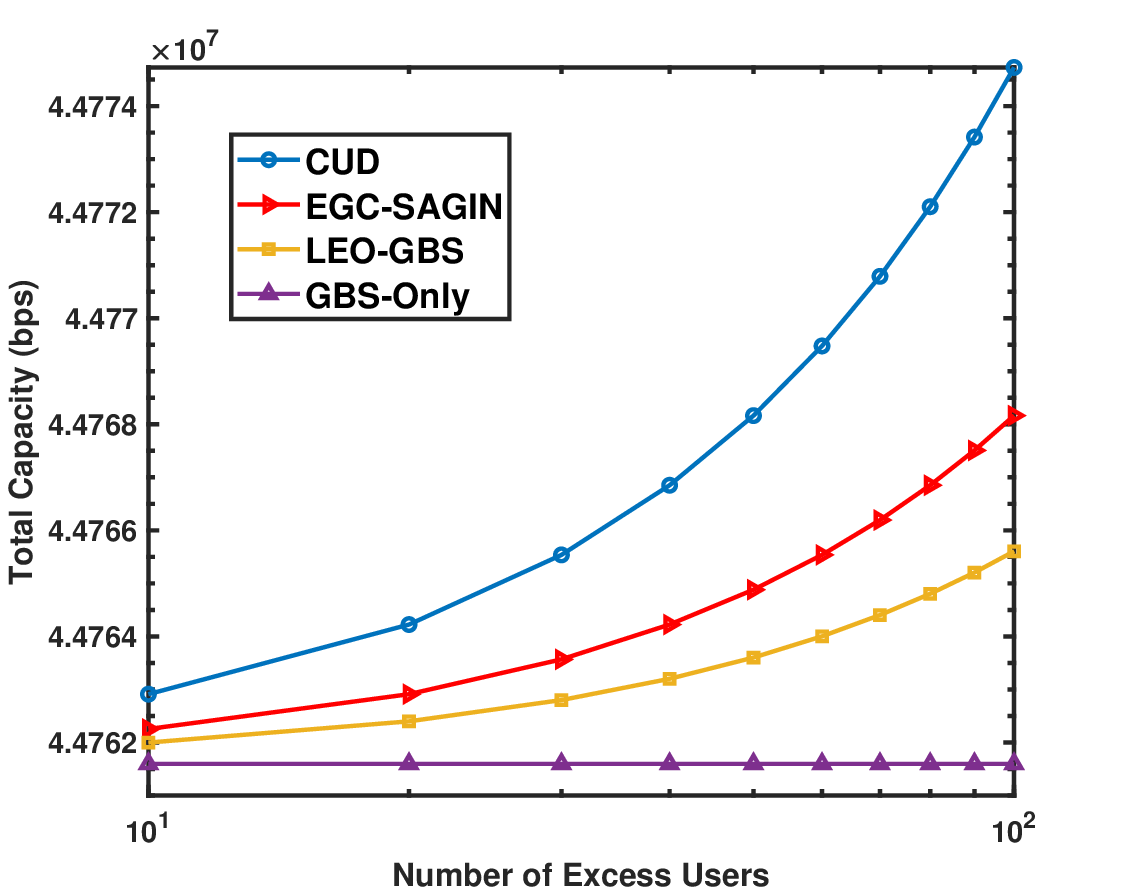}} 
    \subfloat[]{
    \includegraphics[width = 0.25\textwidth]{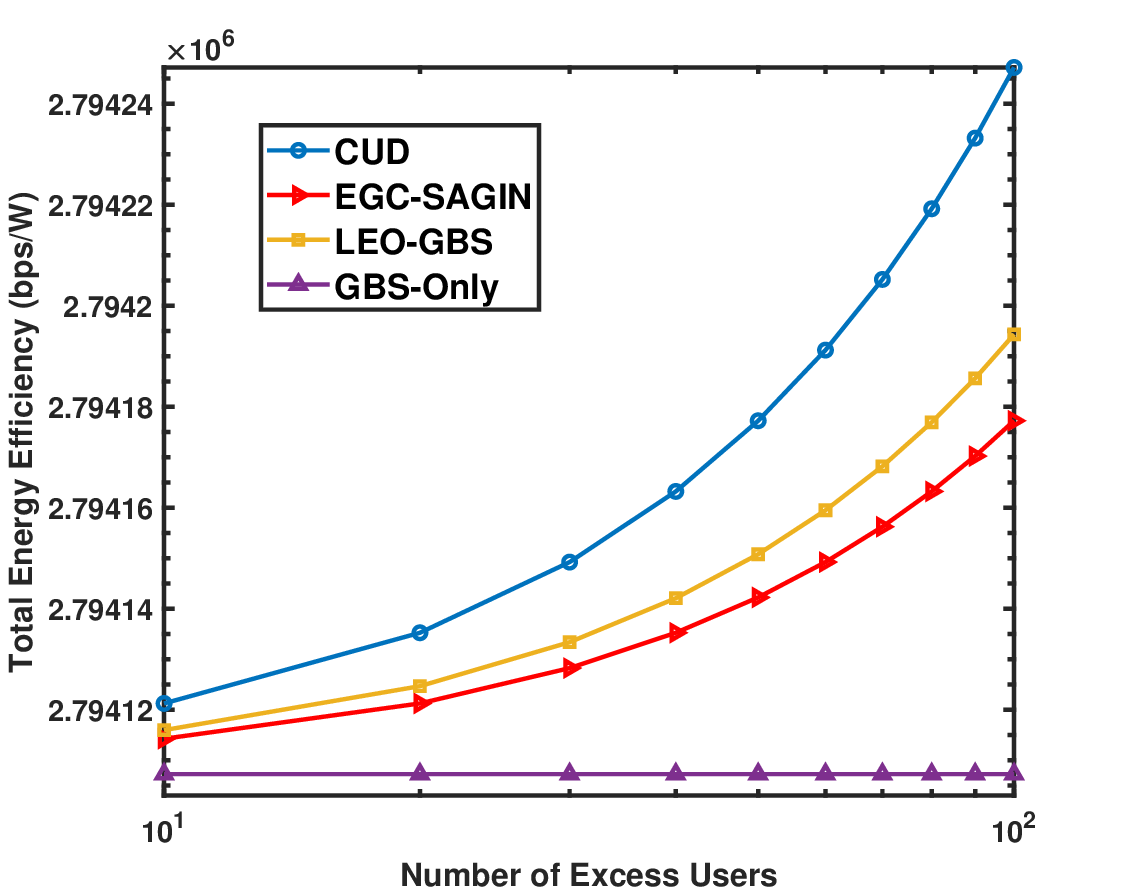}}
    \vspace{-2mm}
    \caption{Illustrating the variation of (a) network capacity, (b) energy efficiency with the number of users.}
    \label{fig:fig4}
\end{figure}
       

%

In Fig.2(b), we compare the performance of the proposed CUD with existing frameworks across varying excess user counts. The CUD consistently outperforms other setups in total network energy efficiency, achieving gains of up to $0.0058$\% ($140$ bps/w) over conventional GBS-only. Compared to EGC-SAGIN and LEO-GBS frameworks, the CUD exhibits improvements of up to $0.0029$\% ($70$ bps/w) and $0.0025$\% ($50$ bps/w), respectively. These results highlight the significant enhancements in energy efficiency enabled by the CUD strategic UAVr deployment and CD-based signal combining.
%


%
%

\section{ Conclusion } 
\label{Sec: Conclusion}
The paper introduces a cooperative signal combining-based UAV relay deployment strategy in satellite air-ground integrated networks. This approach leverages the adaptive deployment of a UAVr in an AF system to enhance the SNR at the user. The cooperative unmanned aerial vehicle relay-based deployment, CUD, framework strategically deploys UAVr in response to fluctuating user traffic, optimizing the user SNR through the cooperative diversity technique. By fostering intelligent and cooperative communication between LEO satellite-user and UAVr-user links, the proposed CUD framework demonstrates significant enhancements in network capacity and energy efficiency at higher UAVr densities. The study highlights the potential of integrating aerial UAVr and LEO satellite-based technologies in future urban communication networks.
%
\section*{Acknowledgment}

This work has been partially funded by the National Science and Technology Council under the Grants NSTC 111-2221-E-A49-071-MY3, and NSTC 113-2221-E-A49 -110 - MY3, and NSTC 112-2634-F-A49-007, and NSTC 113-2218-E-A49-027, Taiwan. This work was supported by the Higher Education Sprout Project of the National Yang Ming Chiao Tung University and Ministry of Education (MOE), Taiwan.

%
\bibliographystyle{IEEEtran}
\bibliography{reference}

\end{document}